\documentclass[preprint,aps,showpacs]{revtex4}
\usepackage{graphicx}
\usepackage{bm}
\usepackage{amsmath}
\begin{document}

\title{Quantum Hamilton-Jacobi Equation and Broken Symmetry in Hydrodynamics of
Liquid Helium II}

\author{S. J. Han}

\affiliation{P.O. Box 4671, Los Alamos, NM 87544-4671}

\begin{abstract}

Based on the quantum theory of Bohm and the phase coherence along
with the mean field of Penrose and Onsager, it is shown that a
free surface of He II behaves like a classical fluid. The broken
symmetry of a macroscopic Bose system at the free surface in an
external field is discussed in terms of the quantum
fluctuations-dissipation. First, we apply this peculiarly
universal behavior to explain a breakdown of superfluidity at a
vortex core. Secondly, we resolve a long standing puzzle with
Landau's two-fluid model on a free surface of a rotating He II in
a gravitational field.

\end{abstract}


\maketitle

The quantization of circulation in He II was suggested over 50
years ago by Onsager and Feynman \cite{Onsager49,Feynman55}. This
prediction was confirmed for irrotational flow in multiply
connected regions by Vinen \cite{Vinen61} who has shown evidence
for the quantization of circulation in units of $h/M$. A further
detailed study of vortex dynamics by Rayfield and Reif using ions
as microscopic probe particles \cite{Reif64} has revealed that,
apart from the quantization of circulation on a macroscopic scale,
the vortices behave like a normal fluid. Yet even now we know little about
the core structure and the immediate region over which a study of
collective excitations may provide clear evidence for the
breakdown of superfluidity; it is, however, exceedingly difficult
to observe an excitation spectrum in such a region which is
smaller than the macroscopic scale of physical measurements. An
extended study of the core structure was undertaken experimentally by
Glaberson, Strayer, and Donnelly in 1968 \cite{Donnelly68,Donnelly95}.
Their result showed that near the vortex line there is a high concentration
of rotons and the \textit {breakdown of superfluidity} at the core as
observed earlier by Rayfield and Reif \cite{Reif64}.

Our initial motivation for studying the vortex dynamics was to
investigate the similarity between the nodal surface of a vortex
core and that of rotating He II in a gravitational field. In the
course of our study, it has become clear that one must address the
question of whether there is any basic underlying mechanism that
could be responsible for both the breakdown of superfluidity and
the parabolic shape of rotating He II. More specifically, it is the
purpose of this paper to show that the basic mechanism is the
spontaneously broken gauge symmetry on the nodal surface. It is shown
here for the first time that the unusual properties of a vortex core are
closely related to those of a free surface of rotating He II.

The study of the free surface of rotating liquid He II began in
1950 by Osborne \cite{Osborne50} to answer the question if the
superfluid component in the two-fluid model \cite{Landau41}
rotates with the normal component. He concluded that, at the
rotational speed used, the superfluid came to a steady rotational
state and the observed curvature of a rotating He II is incompatible
with the two-fluid model.

It is generally accepted that the macroscopic rotation of He II
can be achieved despite the condition $\bm{\nabla}\times
\bm{v}_{s}=0$ if an array of vortices is aligned along the axis of
rotation ($\Omega$) and satisfies Feynman's criteria for the areal
vortex number density $\Gamma_{\Omega}\geq\frac{2\Omega}{\kappa}$
with $\kappa=\frac{h}{M}=0.997\times10^{-3}cm^{2}/sec$ \cite{vortex1}.
We refer the reader Ref.~\cite{Hall60} for details and the counter points of
view in Refs. [9,10].

However, with further experimental evidence by Meservey \cite
{Meservey64} along the same lines as Osborne \cite{Osborne50} in
1964, a question was raised if the theoretical model of the vortex
lines \cite{Hall60} is correct for the interpretation of Osborne's
observation \cite{Osborne50}. Meservey \cite{Meservey64} has also
made a speculation on the role of a free surface energy associated
with velocity discontinuities suggested by Mott
\cite{Mott49}. Such a free surface energy corresponding to
the energy removed by a fixed number of vortices from
the superfluid flow in a manner similar to that of the
Meissner effect in a superconductor could explain how the free surface
rotates in spite of the irrotational motion of He II.

However, the classical fluid like character of the free surface of
rotating liquid helium II still remains as one of the most
puzzling, unanswered questions in low temperature physics since
the first experiment by Osborne \cite{Osborne50}. For a reason
that we shall see in a moment the peculiarly universal property of
the free surface of a superfluid in motion in the gravitational
field can be explained by broken local symmetry; it will be shown
here that, to the approximation to which interactions between
pairs of atoms are considered ({\it i.e.,} the hard sphere
approximation), Meservey's observation \cite{Meservey64} and his
arguments are in agreement with our analysis.

An important step in the development of the vortex dynamics in He
II was the introduction by Anderson of a concept of spontaneously
broken gauge symmetry by which vortices in a superfluid are
nucleated and leads to dissipation by $2\pi$-phase slippage at the walls
\cite{Anderson66,Anderson84}. He further emphasized the main feature
of a flow dissipation mechanism that, at the critical velocity $v_{c}$,
which is much smaller than the Landau's critical velocity
$v_{c}\approx 60m/sec$, a superfluid is unstable to a small perturbation
and this instability drives the motion of a vortex line across orifice,
crossing all the enclosed stream lines \cite{Anderson66}. In such a process a
vortex changes the phase by $2\pi$ and a fixed amount of energy is removed
from the flow - the fundamental flow dissipation mechanism which leads a
breakdown of superfluidity. In an experiment by Avenel and Varoquax
\cite{Avenel85} Anderson's phase-slip picture has been confirmed in a
sub-micron orifice. Since a number of properties of He II and
superconductors can be explained by his theory of broken symmetry,
Anderson's idea has great appeal.

Recently Su and Suzuki \cite{Suzuki01} have shown a new derivation
of the quantization of a vortex, and the similarity between the
rotating He II and the Meissner effect in a superconductor based
on the concept of off-diagonal long-range order (ODLRO) by Penrose
and Onsager \cite{Penrose56} and on the notion of a regauged space
translation. When a particle displacement is introduced into the
Bose system, it invariably perturbs the density of the system.
Moreover, the gauge field is a dynamical variable which must be
analytic unless it is spontaneously broken. We therefore find it
difficult to believe the idea of the regauged space translation
\cite{Suzuki01} is tenable. Moreover this idea led to the
self-contradiction in their work as pointed out by Shi
\cite{Shi03}.

The purpose of the present paper is to show how Bohm's quantum
theory \cite{Bohm52,Aharonov63} can be applied to specific problems in such a
way as to incorporate Feynman's atomic theory of the two-fluid
model \cite{Feynman53} and the concept of the phase coherence
along with ODLRO. By symmetry breaking perturbations in Lagrangian
coordinates \cite{BFKK58,Weinberg67,Han82,Han91}, we study the
consequence of symmetry breaking at the free surface, and thereby
extend Anderson's idea of a symmetry breaking in He II. It will be
shown here that the semiclassical method of symmetry breaking
perturbations in Lagrangian coordinates is in fact a perfectly
well-defined, accurate quantum mechanical approximation scheme.
The unique feature of this perturbation method is that with Bohm's
quantum theory \cite{Bohm52} one is able to separate a surface
phenomenon from that of the bulk fluid.

More specifically, we will first show that the quantization of
circulation is due to the spontaneously broken local symmetry that
accompanies the excitations of highly concentrated rotons at the
core of a vortex as observed in the experiments
and that leads the breakdown of superfluidity at
the vortex line as observed in the experiments by Glaberson,
Strayer and Donnelly \cite{Reif64,Donnelly68,Donnelly95}.

Second, it is shown here that the contour of a free surface of
rotating He II is a necessary consequence of broken symmetry,
which explains why the curvature of the free surface of a rotating
He II remains parabolic \cite{Osborne50,Meservey64}. We also
discuss the similarity between the exclusion of vortices from the
bulk superfluid in a rotating He II below the critical velocity
$\Omega_{c}$ and the Meissner effect in a superconductor.

We begin with the concept of ODLRO in which the reduced density
matrix of the condensed Bose system can be factorized,
\begin{equation}
\rho(r,r^{\prime})=\psi^{\dag}(r)\psi(r)+ \gamma(|r-r^{\prime}|
\label{Odlro}.
\end{equation}
where $\gamma\rightarrow 0$ as $|r-r^{\prime}| \rightarrow
\infty$. The single particle wave function $\psi(r)$ represents
the condensed state in ODLRO and is viewed as macroscopic
dynamical variables. With the the hard sphere approximation of the
inter-particle potential for Bose particles, one can show that the
mean field satisfies the nonlinear Schr\"{o}dinger equation
(Gross-Pitaevskii),
\begin{equation}
i\hbar\frac{\partial\psi}{\partial t}=-\\
\frac{\hbar^{2}}{2M}\nabla^{2}\psi + [V(\bm{x})_{ext} + g_{1}|\psi|^{2}] \psi,\\
\label{Cat2}
\end{equation}
where in the hard sphere approximation $g_{1}=4\pi\hbar^{2}a/M$
and $a$ is the s-wave scattering length \cite{Lee57,note1}.

In our model of an imperfect Bose gas for a superfluid, between
each pair of nearest particles there is a hard sphere repulsion of
range $a$ and no other interaction. This pair-interaction brings
about the Bose condensed state also ensures that the system
possesses the longitudinal collective excitations (phonons). It is
a self-consistent Hartree equation for the Bose condensed wave
function. The whole meaning of an equation such as
Eq.~(\ref{Cat2}) depends on the physical meaning attached to
$\psi(r,t)$, our definition of a superfluid in which the quantum
field $\psi(r,t)$ is a function of macroscopic dynamical variables
which represent a symmetry breaking due to the quantum
fluctuations. Therefore, the whole problem of superfluid dynamics
reduces to the question of how to treat $\psi(r,t)$. To introduce
symmetry breaking perturbations in the atomic theory of two-fluid
model \cite{Feynman53}, we employ Bohm's interpretation of quantum
theory which is essential to our discussion of both phase
coherence and a symmetry breaking in He II.

The essence of Bohm's theory \cite{Bohm52,Aharonov63} is that we
may write the wave function in the form
$\psi(r,t)=f(\bm{r},t)exp[\frac{i}{\hbar}S(r,t)]$, where $S(r,t)$
is an action (or phase) and $\rho=|f(r,t)|^{2}$ may be interpreted
as a number density for a system of $N$ Bose particles.

We then rewrite Eq.~(\ref{Cat2}) to obtain,
\begin{subequations}
\label{allequations} \label{Mean}
\begin{equation}
\frac{\partial\rho}{\partial t}+\bm{\nabla}\cdot(\rho\frac{\bm\nabla S}{M})=0\\
\label{subeq:1}
\end{equation}
\begin{eqnarray}
 \frac{\partial S}{\partial t} + \frac{(\bm\nabla S)^{2}}{2M} +
 V(\bm{x}) - \frac{\hbar^{2}}{4M}[\frac{\nabla^{2}\rho}{\rho}-
\frac{1}{2}\frac{(\nabla\rho)^{2}}{\rho^{2}}]=0, \label{subeq:2}
\end{eqnarray}
\end{subequations}
where $V(\bm{x})\equiv V_{ext} + g_{1}\rho$. Here $ g_{1}\rho$
term represents merely an internal order parameter in the sense of
ODLRO of order-disorder systems \cite{Anderson66} and can be
treated as a pair-interaction potential \cite{Feynman53}.

The last term in Eq.~(\ref{subeq:2}) is the effective quantum
mechanical potential (EQMP) defined by

\begin {equation}
U_{eqmp}=-\frac{\hbar^{2}}{4M}[\frac{\nabla^{2}\rho}{\rho}-
\frac{1}{2}\frac{(\nabla\rho)^{2}}{\rho^{2}}]=
-\frac{\hbar^{2}}{M}\frac{\nabla^{2}f}{f}. \label{Eqmp}
\end{equation}

In the limit $\hbar \rightarrow 0$ (or less restrictively
${\bm\nabla}\rho=0$ for a homogeneous medium), $U_{eqmp}=0$ and
$S(r,t)$ is a solution of the Hamilton-Jacobi equation
\cite{Goldstein}. Eq.~(\ref{subeq:2}) implies, however, that the
particle moves under the action of the force that is not entirely
derivable from the potential $V(\bm{x})$, but which also obtains a
contribution from EQMP. The number density $\rho$ in
Eq.~(\ref{subeq:1}) is not entirely independent, but depends on
$S(\bm{x})$. The solution of Eq.~(\ref{subeq:2}) defines an
ensemble of possible trajectories of a particle in the equation.
We can obtain, in principle, the solution by integrating $v(\bm
x)= {\bm\nabla}S(\bm x)/M$. This is the basic idea of Bohm's
interpretation of quantum theory \cite{note2}. Finally,
Eq.~(\ref{subeq:2}) also suggests us the way of determining the
most stable motion about which to linearize the dynamical
equations, Eqs.~(\ref{Mean}).

Let us now consider to what degree the phase coherence can be
defined in ODLRO for a many-body problem. Suppose if we set up a
state in which a small number of atoms occupy in a ring around the
vortex line and each atom is displaced by an infinitesimal
position perturbation as the Feynman model suggests
\cite{Feynman55,Hall60}, the phase coherence can be defined as a
correlation between an individual particle phase change and that
of the mean field by introducing perturbations in Lagrangian
coordinates. The position perturbation of a particle is given by
$\bm{x}_{i}=\bm{x}_{0,i} +\bm{\xi}(\bm{x}_{0,i},t)$, where
$\bm{\xi}$ is a function of the initial position of a particle and
time in the many-body wave function for $n$-particles in the ring,
and remains attached to the particle in motion
\cite{BFKK58,Weinberg67,Han82}.

In order to show the phase coherence in a many-body problem in the
ring, we introduce the wave function,
$\psi(\bm{x}_{1},\bm{x}_{2},\cdots,\bm{x}_{n})=f(\bm{x}_{1},\bm{x}_{2},
\cdots
\bm{x}_{n})\exp[iS(\bm{x}_{1},\bm{x}_{2},\cdots,\bm{x}_{n})]$,
where $n$ is the number of particles. $f^{2}$ is equal to the
density of representative points
$(\bm{x}_{1},\bm{x}_{2},\cdots,\bm{x}_{n})$ in $3n$-dimensional
ensemble in a Bose system. To the first-order in $\bm{\xi}$,
\begin{equation}
S_{i}(\bm{x}_{1},\bm{x}_{2},\cdots,\bm{x}_{n})=S_{i}(\bm{x}_{0,1},
\bm{x}_{0,2},\cdots, \bm{x}_{0.n}) +
\bm{\xi}_{i}\cdot\bm{\nabla}_{0,i}S_{i}(\bm{x}_{0,1},\bm{x}_{0,2}, \cdots,
\bm{x}_{0,n}).
\label{phase1}
\end{equation}

With the Lagrangian displacement vector
$\bm{r}=\bm{r_{0}}+\bm{\xi}(\bm{r}_{0},t)$ which is now a position
vector of a particle described by the one-particle wave function
associated with the Bose condensed wave function (mean field) in
the reduced density matrix in ODLRO, we may calculate the
first-order in $S$, $\bm{v}$ and $\rho$ for the mean field
\cite{BFKK58,Weinberg67,Han82},
\begin{subequations}
\label{allequations} \label{First}
\begin{eqnarray}
S(\bm{r},t)=S(\bm{r}_{0},t) +
\bm{\xi}\cdot\bm{\nabla}_{0}S(\bm{r}_{0},t)\\
\label{Firsta} \bm{v}(\bm {r},t)=\bm {v}(\bm {r}_{0},t) + \frac
{\partial}{\partial t}\bm{\xi} +\bm{v}(\bm{r}_{0},t)\cdot
\bm{\nabla}_{0}\bm{\xi}, \label{Firstb}\\
\rho(\bm{r},t)=\rho(\bm{r}_{0})-
\bm{\nabla}_{0}\cdot[\rho(\bm{r}_{0})\bm{\xi}] \label{Firstc},
\end{eqnarray}
\end{subequations}
where $\bm{\nabla}_{0}$ denotes the partial derivative with
respect to $\bm{r}_{0}$ with $\bm{\nabla}\rightarrow
\bm{\nabla}_{0}-\bm{\nabla}_{0}\bm{\xi}\cdot\bm{\nabla}_{0}$. And
$\bm{v}(\bm{r}_{0},t)=\bm{\nabla}S(\bm{r}_{0},t)/M$ is a solution
of the Hamilton-Jacobi equation and $S(\bm{r}_{0},t)$ becomes a
phase \cite{Bohm52,Aharonov63}. Eq.~(\ref{Firstc}) was obtained
from Eq.~(\ref{subeq:1}) by using Eq.~(\ref{Firstb}) and then
integrating over the time \cite{Weinberg67}.

The first-order expansions in Eq.~(\ref{First}) are semiclassical
since $S$ is a phase \cite{Aharonov63} and $\bm{\xi}$ is is a
classical variable in a Lagrangian coordinate. The semiclassical
perturbation method to the solution of the Hamilton-Jacobi
equation is in fact a well defined, accurate quantum mechanical
approximation scheme provided that $\psi(\bm{x},t)$ satisfies the
nonlinear Schr\"{o}dinger equation Eq.~(\ref{Cat2}). Later we
shall show this by deriving the Bogoliubov spectrum \cite{Bog47},
but in the meantime we shall describe the scheme from
semiclassical point of view.

It is useful to define \textit {the boundary conditions} in terms
of $\bm{\xi}$ at a free surface, {\i.e.,}
$\bm{\nabla}\cdot\bm{\xi}=0$ and $\bm{\nabla}\times\bm{\xi}=0$,
which follow from Eq.~(\ref{First}) by the conditions of
incompressibility and irrotational motion of a fluid at the free
surface \cite{SJHan1,Lamb45,Landau59}.

Our analysis of the quantization of circulation in He II requires
the introduction of the phase coherence which is defined in ODLRO:
$\bm{\xi}\cdot\bm{\nabla}S(\bm{x}_{0},t)
=\sum_{i}\bm{\xi}_{i}\cdot\bm{\bm{\nabla}_{i}}S_{0,i}(\bm{x}_{0,i},t)$,
where the summation of the phase change by an individual atom on
the right-hand side is not measurable, but the left-hand side in
the mean-field is the one that can be measured in an experiment
\cite{Aharonov63,Inouye01}. Because of the phase coherence, the
motion of one atom also implies the motion of others along the
nodal surface. It must also be emphasized here that since the
action $S$ is a macroscopic dynamic variable which represents
symmetry breaking, it would, as in the case of electromagnetic
field, depend to some extent on the actual location of the atom,
and the perturbed terms become important in a physical process
involving short distances (in $\AA$).

Unlike the elementary excitations of quasi-particles (phonons and
rotons) in He II, the macroscopic excitation of vortices takes
place in a flow of a large amount of fluid at considerably higher
energy \cite{Reif64,Vinen63}. We follow closely Feynman's picture
of quantization of circulation \cite{Feynman55,Hall60} and
consider a ring of atoms that are located along the nodal surface
for which the boundary conditions, $\bm\nabla\times\bm\xi=0$ and
$\bm\nabla\cdot\bm\xi=0$ must be satisfied, and that the atoms are
rotating under the gravitational field. The effect on the wave
function is then given in the form, equivalent to
Eq.~(\ref{phase1}). In ODLRO, the mean field $\psi(\bm{x},t)$ is a
single-valued function, but the phase $S$ need not be
single-valued in a multiply-connected system \cite{Anderson66}; it
need merely to return to its original value by traversing around a
nodal surface.

With the spontaneously broken gauge symmetry, the Onsager-Feynman
quantization \cite{Onsager49,Feynman55,Hall60} can be stated in a
more precise manner,
\begin{equation}
\sum_{cir} \bm{\nabla}S\cdot\bm{\xi}= \oint\bm{\nabla}S\cdot
d\bm{l}= nh. \label{vortex}
\end{equation}
Here the summation is taken around the nodal surface and
$\bm{\xi}$ is taken to be small since it is an atomic
displacement. In He II, if one takes $\bm{\nabla}S(\bm{x}_{0},t)=
M \bm{v}(\bm{x}_{0})_{s}$, then $\kappa=\oint\bm{v}_{s}\cdot
d\bm{l}$, where $\kappa=h/M=0.997\times10^{-3}cm^{2}/sec$.

One sees that the spontaneously broken gauge symmetry is more
fundamental for the creation of a vortex than that of the Bohr-Sommerfield
condition. The fact that the experimental data fit almost perfectly
on the curve of a relation between the velocity $v$ and energy $E$
of a vortex ring  with $\kappa=h/M$ and the core radius $a=(1.28\pm 0.13)\AA$
explains why the vortices behave entirely classically, since except for $\kappa$
value both $v$ and $E$ are derived from classical hydrodynamics. It is therefore
clear that the broken gauge symmetry leads to a breakdown of superfluidity.

The most definite demonstration of this concept was an observation of phase change
by Inouye, {\it et al.,} \cite{Inouye01}. Moreover, it shows that
$\bm{\nabla}\times\bm{\nabla}S=0$ as it must be unless the gauge
symmetry is spontaneously broken
\cite{Anderson62,Anderson84,BenLee73,Weinberg96}.

There is no simple mathematical prescription for demonstrating
that the symmetry of a Bose system is broken at a free surface of
He II in a gravitational field. Therefore in order to demonstrate
that the symmetry be always broken at a free surface of He II, it
is necessary to introduce the collective excitations that take
place simultaneously on both sides of the boundary for the
solutions to the boundary-value problems in superfluid dynamics.

As a simple model which retains the essential features of the
problem, we consider the free surface of a uniformly flowing
superfluid. First, we consider the response of a free surface in a
uniform superfluid flow to perturbations as a model for a reason
that will become apparent shortly. Let us consider the superfluid
in motion in the z-plane under the gravitational field. Here we
assume the amplitude of a surface wave is small compared to the
wavelength and thus take for the surface wave as the potential
flow $\nabla^{2}\phi=0$ with $\bm{\xi}_{s}=-\bm{\nabla}\phi$,
which follows from the boundary conditions
$\bm{\nabla}\cdot\bm{\xi}=0$ and $\bm{\nabla}\times\bm{\xi}=0$
\cite{SJHan1,Lamb45,Landau59}. In this connection, it is important
to recall that the use of the Hamilton-Jacobi equation in solving for
the motion of a particle is only a matter of convenience. Hence we employ
Euler's equation (with $\nabla p=0$) to describe the surface waves:

\begin{equation}
\rho(\frac{\partial}{\partial t}\bm{v}+\bm{v}\cdot \bm{\nabla} \bm
{v})=-\rho\bm{g}\cdot\bm{\nabla}(\bm{r}), \label{Motion0}
\end{equation}
where $\bm{g}$ is the gravitational acceleration.

In the rest frame of a fluid by Galilean transformation, we may take
$\rho_{0}=\rho_{0}\Theta(-z)$ and $\bm{r}=\bm{r}_{0}+\bm{\xi}$ and linearize
Eq.~(\ref{Motion0}) along with $\bm{\xi}_{s}=-\bm{\nabla}\phi$ as a displacement
vector, and Eqs.~(\ref{Firstb})-(\ref{Firstc}). After a brief algebra we obtain
the well-known dispersion relation for the gravity wave:

\begin{equation}
\omega^{2}=g\,k,
\label{Swave1}
\end{equation}
where $\phi=A exp[kz]cos(kx-\omega t)$ is assumed and k is the wave number.

To incorporate the capillary wave as a part of the surface wave, we must
modify the potential $\phi$ which is a solution of the Poisson equation.
Furthermore, we note that the equation of motion for a particle at the free
surface of the fluid is quite different in that the pressure difference between
two sides of the free surface is not zero ({\it i.e.,} $\bm{\nabla} p\neq0$
in Euler's equation) and the potential $\phi$ should instead satisfy the
Laplace formula \cite{Landau59,Sommerfeld50},

\begin{equation}
\rho g\frac{\partial\phi}{\partial z}+\rho\frac{\partial^{2}\phi} {\partial
t^{2}}-\alpha\frac{\partial}{\partial z} [\frac{\partial^{2}\phi}{\partial
x^{2}}+\frac{\partial^{2}\phi}{\partial
 y^{2}}]=0,
\end{equation}
where $g$ is the gravitational acceleration and $\alpha$ is the
surface tension, $\alpha\simeq 0.34 erg/cm^{2}$ for He II. If we
assume $\phi=A\exp[kz]cos(kx-\omega t)$, we are led immediately to
the dispersion relation for the gravity wave,
\begin{equation}
\omega^{2}=gk +\frac{\alpha k^{3}}{{\rho}_{s}} \label{Swave2}.
\end{equation}

The meaning of this dispersion relation is obvious: in the long
wavelength limit, $k \ll (g\rho/\alpha)^{1/2}$, we
have a pure gravity wave; in the short wavelength limit ({\it
i.e.,} in $\AA$), the capillary waves obey the dispersion relation
$\omega^{2}=\frac{\alpha k^{3}}{{\rho}_{s}}$ with the density
$\rho_{0}$ at the free surface; the capillary waves have been
observed in a recent experiment by Elliott, {\it et al.,}
\cite{Elliott00}.

The most remarkable feature of Eqs.~(\ref{Swave1}) and (\ref{Swave2}
is that the dispersion relation is independent of internal dynamics of a
superfluid, {\it i.e.,} it is independent of the pair-interaction
potential, and depends only on the external gravitational
potential and the surface tension. The free surface thus behaves
like a classical fluid, because it is independent of both $\hbar$ and
the speed of first sound
$c=[\frac{4\pi a \rho_{0}\hbar^{2}}{M^{2}}]^{1/2}$ \cite{Bog47}.

With this tacit understanding, which will play an important role
later, we shall next proceed with the calculation of the
excitation spectrum for phonons in the fluid. Indeed, we have discovered
this universal property of a free surface during the course of our study of
collective excitations in a Bose condensate in trap in which the trapping force
is comparable to the pair-interaction force of Bose particles in the system.

Inside of the free surface, the fluid is compressible,
$\bm{\nabla}\cdot\bm{\xi}\neq0$, but it remains irrotational
$\bm{\nabla}\times\bm{\xi}=0$ as it is a superfluid, which follows
from Eq.~(\ref{First}). To obtain the phonon excitation spectrum,
first note that the vortices are in an isolated region, so that we
assume a uniform superfluid flow $\bm{v}(\bm{x}_{0})$, and take
the first-order terms $\rho_{1}$ and $S_{1}$ varying as
$C\exp[i(\bm{k}\cdot\bm{x}-\omega t]$. It is then straightforward
algebra to obtain the first-order linearized equations from
Eq.~(\ref{First}),
\begin{subequations}
\label{whole}
\begin{eqnarray}
-i\omega\rho_{1}-\frac{\rho_{0}}{M}k^{2}S_{1}=0 \\
\label{lasta} -i\omega S_{1}+\frac{4\pi \hbar^{2} a}{M}\rho_{1}
+\frac{\hbar^{2}}{4M}\frac{1}{\rho_{0}}k^{2}\rho_{1}=0.
\label{lastb}
\end{eqnarray}
\end{subequations}

Eqs.~({\ref{whole}) yield at once the dispersion relation:
\begin{equation}
\omega^{2}=\frac{4\pi a\rho_{0}\hbar^{2}k^{2}}{M^{2}}+
\frac{\hbar^{2}k^{4}}{4M^{2}} \label{Bog47}
\end{equation}

This is the well-known Bogoliubov spectrum in He II \cite{Bog47}.
It is evident why our semiclassical perturbation method is
actually a perfectly well defined quantum mechanical approximation
scheme.

We comment here briefly that, in the phonon regime $k\rightarrow
0$, $\omega=[\frac{4\pi a \rho_{0}\hbar^{2}}{M^{2}}]^{1/2} k$. The
energy spectrum is characteristic of a sound wave with the speed
of first sound $c=[\frac{4\pi a \rho_{0}\hbar^{2}}{M^{2}}]^{1/2}$
in an imperfect Bose gas. It is therefore natural to identify the
underlying basic mechanism for the phenomena as \textit{a
spontaneously broken gauge symmetry} at the free surface in a Bose
system ({\it i.e.,} a transition from a superfluid to a normal
fluid). The phonons may be interpreted as the massless
Nambu-Goldstone bosons as $k\rightarrow 0$ in the symmetry
breaking \cite{Anderson62,BenLee73,Weinberg96}. It should be
emphasized that since in reality phonons are never confined to a
strictly mathematical surface, the surface layer is thus composed
of phonons and particles, which is consistent with the property of
a free surface.

In connection with the two entirely different waves in He II under the
gravitational field, it is worth while to note that Eq.~(\ref{Cat2})
is the second-order partial differential equation. When it is linearized
for an isolated system, it yields the second-order partial differential
equation in $\bm{\xi}$ which gives two particular solutions: one being
the surface wave, the other the sound wave \cite{Han91,Han05}. In a linear
regime, each wave satisfies its own dispersion relation. As shown above,
the surface waves are always independent of internal dynamics - a classical
fluid like behavior \cite{Han05}. If in fact the gauge symmetry is not broken,
we cannot hold the law of conservation of total energy in an isolated system
\cite{Han05}.

The basic mechanism of symmetry breaking can be described as the
following: as the sound wave propagates, it loses its phase
coherence due to quantum fluctuations of particle
trajectories driven by the effective quantum mechanical potential
$U_{eqmp}$ and hence the sound wave dissipates by the interaction
with a normal fluid at the free surface, which is also an irreversible
process as the fluctuation-dissipation theory of Kubo
\cite{Kubo57}. It is more fundamental than that of Kubo's theory of
fluctuation-dissipation although both approaches are based on the
dynamical variables (hidden variables in Bohm's theory), since the
fluctuation-dissipation is due to the broken gauge symmetry and the law of
conservation of total energy in an isolated system can be preserved as a
consequence of the broken gauge symmetry.

It should be stressed that the dissipation process gives rise the surface
energy \cite{Elliott00} which in turn raises the energy of
particles in the surface layer; and therefore that the particles
in the layer obey the Boltzman statistics just like rotons in
Landau's two fluid model - a Bose quantum fluid to a normal fluid.
Thus, we see that a complete picture of the two-fluid model
emerges from this analysis \cite{Landau49}.

The above analysis [Eq.~(\ref{Swave1}) and Eq.~(\ref{Bog47})] is
useful in the discussion of symmetry breaking at a free surface of
a superfluid under the gravitational field [{\it i.e,} a breakdown
of superfluidity at a vortex core and at the free surface of
rotating He II]. It says two things: first, that the symmetry of a
Bose system under an external field is broken for all wave numbers
at the free surface. It is broken regardless of a local curvature
of the free surface, for example, $k_{\theta}=m/r$ in cylindrical
geometry, where $m$ is the mode number and $r$ is the radial
distance of a point in cylindrical coordinates \cite{Han82}.
Secondly, the free surface always behaves like a classical fluid -
a breakdown of superfluidity.

The first experimental observation on the collective excitations
of rotons localized near the vortex core was reported in 1968
\cite{Donnelly68}. Keeping in mind the picture of a roton as the
quantum analog of a smoke ring whose excitation energy is much
higher than that of phonons, we limit, for the sake of simplicity,
our discussion on the mechanism that drive roton excitations near
the vortex line.

First we note that the free surface at the vortex core is a nodal
surface \cite{Hall60} and should therefore behave like a normal
fluid. This explains why Rayfield and Reif \cite{Reif64} have
observed that a vortex responds like a normal fluid to an ion
probe at the core. Moreover, as the superfluid density $\rho_{s}$
approaches to zero near the nodal surface of a vortex which
rotates approximately with the speed of the first sound
$c=\kappa/(\sqrt{8}\pi\eta)$ with $\eta$ the coherence length,
EQMP fluctuates rapidly to excite the rotons near the vortex line,
because on the vortex line $U_{eqmp}$ is highly singular.

To show this explicitly, let us take a close look at EQMP. As
emphasized by Bohm \cite{Bohm52}, the particle experiences a force
from EQMP which fluctuates with the ensemble average energy and
its particle momentum $\bm{p}=\bm{\nabla}S$ ({\it e. g.,}
$v_{s}=\kappa/2\pi r$ for a quantized vortex). If a particle
happens to enter a region of space where $\rho$ becomes small,
these fluctuations can become exceedingly large. Even with the
vortex core model $\rho=\rho_{0}\frac{r^{2}}{r^{2}+\eta^{2}}$
\cite{Clem75}, EQMP fluctuates with the degree of divergence
$\frac{\hbar^{2}}{2 \pi }\frac{1}{r^{2}}$ as $r\rightarrow 0$. It
is also obvious that the degree of fluctuations depends on the
core radius and thus the local radius of a curvature
\cite{Hess95}. Hence it is apparent that the concentrated rotons
near the vortex core are due to the spontaneously broken gauge
symmetry that accompanies  rotons (pseudo-Goldstone Bosons)
\cite{BenLee73,Weinberg96}; the free (nodal) surface behaves like
an ordinary fluid - normal fluid.

Just as in the case of the free surface He II under the
gravitational field, the particles in the nodal surface of a
vortex line obey the Boltzman statistics due to the scattering
with concentrated rotons; a remarkably simple picture which
explains why Vinen's observation of a vortex quantization based on
the classical Magnus force remains correct \cite{Vinen61}. This
picture, which is also in agreement with the observation of
Rayfield and Reif \cite{Reif64}, resolves the recent controversy
over the Magnus force \cite{Thouless99}. While we have not made
any attempt to derive the roton excitation spectrum
\cite{Feynman54}, we feel our analysis captures the essence of the
broken symmetry \cite{Anderson62,BenLee73,Weinberg96} in a
superfluid.

Because of the peculiarly universal classical behavior of the free
surface, the breakdown of superfluidity observed in the experiment
\cite{Reif64,Donnelly68} can be considered as a manifestation of the
broken symmetry which accompanies pseudo-Goldsone bosons (rotons),
{\it i.e.,} a transition from a superfluid fluid to a normal
fluid. This analysis also confirms Anderson's insight on the role
of quantum fluctuations in the symmetry breaking of the
macroscopic Bose system \cite{Anderson62,Anderson66,Anderson84}.

We finally turn to the question of whether the superfluid
component in a rotating He II does in fact rotate. Osborne's
observation \cite{Osborne50} is believed to be incompatible with
Landau's two-fluid model \cite{Landau41} from which he derived the
following equation for the rotating He II:

\begin{equation}
z=\frac{\rho_{n}}{\rho}\frac{\Omega^{2}}{2g}r^{2},\label{Two-fluid}
\end{equation}

where $\rho=\rho_{s}+\rho_{n}$, $\Omega$ is the angular velocity,
$g$ is the gravitational acceleration, and $z$ and $r$ are the
vertical and radial coordinates of a point on the surface with
respect to the vortex at the origin.

As shown above in a uniform flow of a superfluid, the free surface
is independent of internal dynamics and behaves like a classical
fluid. With this understanding, it is not difficult to derive the
equation for the contour of a free surface of rotating He II in a
steady state under the gravitational field \cite{Sommerfeld50}:

\begin{equation}
z=\frac{\Omega^{2}}{2g}r^{2}.\label{Broken}
\end{equation}

This is the expression confirmed in detail in experiments of a
rotating He II both by Osborne \cite{Osborne50} and by Meservey
\cite{Meservey64}. Here it is derived from the conditions that the
density at the free surface is given by $\rho_{0}$ which follows
from Eq.~(\ref{First}) with the boundary condition at the free
surface. Also in the derivation of the above equation, we treated
the superfluid as compressible ($\bm{\nabla}\cdot\bm{\xi}\neq 0$)
outside the nodal surface ({\it i.e.,} away from $z$-axis, a
vortex line), yet remaining irrotational
($\bm{\nabla}\times\bm{\nabla}S=0$ unless it is spontaneously
broken) to be consistent with the quantum Hamilton-Jacobi
equation. Here we wish to stress that the rotation of the bulk
superfluid is not necessary in this derivation.

As emphasized by Mott, \textit{"the essential point in the
two-fluid model is that when the normal fluid and superfluid are
set in motion relative each other, there is no transfer of
momentum from one to the other"} \cite{Mott49}. The question
arises then as to why the free surface of the superfluid does
rotate with the angular velocity ($\Omega$) of the normal fluid to
form a parabolic surface due to the centrifugal and gravitational
forces on the surface. To answer this question we note that in
reality a surface flow is not confined to a mathematical surface
but to a surface layer with a finite thickness, which also
contains the rotons, phonons, and He atoms with much higher energy
than that of the bulk superfluid. Hence the free surface plays the
role of a wall in Anderson's analysis of a vortex nucleation
\cite{Anderson66,Anderson84}.

Therefore, we can see the vortices created at the solid base of a
cylindrical container diffuse upward to the free surface causing
the rotation of the free surface through the exchange of the
angular momenta of the vortices by $2 \pi$ phase slippage that
also accompanies dissipation process which gives rise to a free
surface energy at the same time. As a result, the free surface
rotates with the angular velocity $\Omega$ with the normal fluid
which was examined by Meservey in his empirical study on the
transient effects \cite{Meservey64}. Another important point is
that the free surface energy manifested in a recent observation of
capillary waves at the free surface of He II \cite{Elliott00} may
indeed exclude vortices from the bulk superfluid in a manner
similar that of the Meissner effect in a superconductor.

Our results are therefore just what we should expect on physical
grounds - a picture of the quantum fluctuation-dissipation
\cite{Kubo57}. This picture emerges as a result of the broken
symmetry and the phase slippage of vortices at the free surface,
which is also consistent with Anderson's insight into the role of
quantum fluctuations on the symmetry breaking
\cite{Anderson62,Anderson66}. More importantly, the present
analysis confirms Meservey's insight into the dynamics of upward
diffusion of vortices from the solid base in a rotating He II
\cite{Meservey64}.

Now it may be asked why this obvious inconsistency in the
parabolic shape of rotating He II has long remained unresolved in
Landau's two-fluid model. To answer this question, it is helpful
to consider Landau's quantum hydrodynamics \cite{Landau41}, in
which he emphasized the superfluid velocity and its equation of
motion in an infinite uniform fluid. Bogoliubov also developed a
microscopic theory of superfluidity in He II using a model
Hamiltonian in quantum field theory \cite{Bog47}, which is valid
only in a Hilbert space. Hence they have not addressed the surface
phenomena. In 1966 Anderson introduced the concept of broken gauge
symmetry in quantum fluids to explain the vortex nucleation
\cite{Anderson62,Anderson66} after Josephson's discovery of
macroscopic interference phenomena \cite{Josephson62}. However,
the phenomena such as breakdown of superfluidity at a vortex core
have remained unsolved in low temperature physics to date.

In summary, it is shown that the symmetry of a Bose system is
broken at a free surface which accompanies quasi-particles
(phonons and rotons) depending on the radius of curvature of the
free surface and the flow velocity. This peculiarly universal
classical fluid like behavior of a free surface in a rotating He
II is a necessary consequence of the broken symmetry. One may thus
interpret the breakdown of superfluidity at the vortex core as the
spontaneously broken gauge symmetry that accompanies the
pseudo-Goldstone bosons (rotons). The semiclassical perturbation
method using the Lagrangian displacement vector is indeed a
perfectly well-defined quantum mechanical approximation scheme. As
envisaged for the microscopic phenomena at short distances ({\it
i.e.,} several $\AA$) in a many-body system by Bohm \cite{Bohm52},
EQMP plays a crucial role in the dynamics of a many-body
interacting system at the quantum level of accuracy. It also
explains why the vortices with an ion probe at the core move like
a normal fluid \cite{Reif64}. We have also shown that the free
surface energy due to surface tension in a rotating He II may
indeed exclude the vortices from the bulk He II just like the
Meissner effect in a superconductor, which explains Meservey's
observation on the contour of rotating He II.

I am grateful to Professor W. F. Vinen for pointing out the role
of surface tension at the free surface of He II. I am also
indebted to Dr. Robert Meservey for his helpful comments on the
upward diffusion of vortices from the solid base in his rotating
He II experiment. I also thank Professor G. B. Hess for sending me
his paper on the geometrical effect of nucleation rate of
vortices.

{\it Note added in proof}. After this paper was written, I learned
about 't Hooft's conjecture [Massimo Blasone, Petr Jizba, Giuseppe
Vitiello, Phys. Lett. A 287 (2001) 205]. The Letter describes the
dissipation process of the 1-D classical oscillators with a
damping term (coupled by time-reversal) by quantization of the
orbits. In this paper, the symmetry of a Bose system is broken by
the quantum fluctuation of the classical orbits defined by
Hamilton-Jacobi equation near the free surface. The picture of
fluctuation-dissipation in this paper is essentially equivalent to
't Hooft's conjecture.

\newpage


\begin{thebibliography}{95}
\bibitem{vortex1} What has been overlooked in the Feynman vortex line model
        is the stability of a vortex line which is unstable to $m=1$
        the hose mode (to be published).
\bibitem{note1} The Gross-Pitaevskii equation with a hard sphere
        approximation for pair-particle ineractions describes a weakly
        interacting Bose system, but not He II which is a srtongly
        interacting Bose system. It is one defect of the present analysis.
        Nevertheless, the model of an imperfect Bose gas explains
        many features of the Landau's two-fluid model. Moreover, we can
        describe the dynamics of the symmetry breaking in He II as
        an anlytically tractable model. 
\bibitem{note2} Because of the presence of $g_{1}\rho$ term in $V(\bm{x})$,
        Bohm's tentative interpretation of $\rho$ as a field coordinate
        and $S(\bm {x})$ as the momentum, canonically conjugate to $\rho$ is no
        longer valid.
\end{thebibliography}
\end{document}